\documentclass[10pt, reqno]{article}
\pdfoutput=1

\usepackage{jheppub}
\usepackage{amssymb}
\usepackage{amsmath}
\usepackage{mathtools}
\usepackage[usenames,dvipsnames]{xcolor}
\usepackage{epsfig}
\usepackage{dcolumn}
\usepackage{esvect}
\usepackage{tikz}
\usetikzlibrary{shapes.geometric, arrows}
\usepackage{upgreek}
\usepackage{setspace}
\usepackage{enumitem}
\usepackage{array,multirow,bigdelim,arydshln}
\usepackage{appendix}
\usepackage{xparse}
\usepackage{empheq}
\usepackage{dsfont}
\usepackage[utf8]{inputenc}
\usepackage{microtype}
\hypersetup{
	colorlinks,
	urlcolor=Maroon,
	linkcolor=Maroon,
	citecolor=Maroon
}

\def \tr {\nonumber\\}
\def \be {\begin{equation}}
\def \en {\end{equation}}
\def \bes {\begin{eqnarray}}
\def \ens {\end{eqnarray}}
\def \vol {\rm vol\,}
\def \slr {{\rm SL}(2, \mathbb{R})}
\def \slc {{\rm SL}(2, \mathbb{C})}
\def \PT {\textsf{PT}}
\def \PTap {\textsf{PT}_{\alpha'}}
\def \I {\mathbb{I}}
\def \CHY {d\mu^{\rm CHY}_n}
\def \KN {d\mu^{\rm KN}_n}
\def \kn {\mathsf{KN}}
\def \Pfp {{\rm Pf}^\prime}

\def \ibp {\overset{\tiny\begin{subarray}{c}\text{IBP}\end{subarray}}{=}}

\title{A String Deformation of the Parke--Taylor Factor}
\author{Sebastian Mizera}
\author{and Guojun Zhang}
\affiliation{Perimeter Institute for Theoretical Physics, Waterloo, ON N2L 2Y5, Canada}
\affiliation{Department of Physics \& Astronomy, University of Waterloo, Waterloo, ON N2L 3G1, Canada}
\emailAdd{smizera and gzhang2@pitp.ca}

\abstract{Scattering amplitudes in a range of quantum field theories can be computed using the Cachazo--He--Yuan (CHY) formalism. In theories with colour ordering, the key ingredient is the so-called Parke--Taylor factor. In this note we give a fully $\text{SL}(2,\mathbb{C})$-covariant definition and study the properties of a new integrand called the \emph{string Parke--Taylor} factor. It has an $\alpha'$ expansion whose leading coefficient is the field-theoretic Parke--Taylor factor. Its main application is that it leads to a CHY formulation of open string tree-level amplitudes. In fact, the definition of the string Parke--Taylor factor was motivated by trying to extend the compact formula for the first $\alpha'$ correction found by He and Zhang, while the main ingredient in its definition is a determinant of a matrix introduced in the context of string theory by Stieberger and Taylor.\vspace{1em}}

\toccontinuoustrue

\begin{document}

\maketitle
\addtocontents{toc}{\protect\setcounter{tocdepth}{1}}
\numberwithin{equation}{section}

\pagebreak

\section{Introduction}

Since the seminal work of Witten \cite{Witten:2003nn} on twistor string theory, numerous methods for calculating field-theory scattering amplitudes utilizing Riemann surfaces have been developed \cite{Witten:2003nn,Roiban:2004yf,Berkovits:2004hg,Cachazo:2012da,Cachazo:2012kg,Cachazo:2012pz,Cachazo:2013gna,Cachazo:2013hca,Cachazo:2013iea,Cachazo:2014nsa,Cachazo:2014xea,Mason:2013sva}. One such prescription, known as the Cachazo--He--Yuan (CHY) formalism \cite{Cachazo:2013gna,Cachazo:2013hca,Cachazo:2013iea,Cachazo:2014nsa,Cachazo:2014xea}, is independent of the space-time dimension and works for a large class of theories.

Much work has been done towards relating the CHY formalism to string theory \cite{Berkovits:2013xba,Mason:2013sva,Gomez:2013wza,Bjerrum-Bohr:2014qwa,Siegel:2015axg,Casali:2016atr,Li:2017emw}. One approach is to think about the field-theory amplitude as the leading-order $\alpha'$-coefficient of a string theory amplitude, and consider finding corrections in its $\alpha'$-expansion. A big step towards this direction has been recently made by He and Zhang \cite{He:2016iqi,Zhang:2016rzb}, who found compact CHY integrands for the sub-leading $\alpha'$-corrections to the open and closed string amplitudes in both superstring and bosonic sectors. Following this approach we can define a CHY formula for the open superstring $n$-pt partial amplitude as Yang--Mills plus string corrections:
\bes\label{intro1}
{\cal A}^{\rm open}(\beta) = \int \CHY\, \underbrace{\Big(\, \PT(\beta) + \ldots\, \Big)}_{\PTap(\beta)}\, \Pfp \Psi,
\ens
where $\CHY\!$ denotes the CHY measure. All the $\alpha'$ and colour dependence is encapsulated in the left half-integrand, while the whole information about the polarization vectors is contained in the right half-integrand. In fact, this decomposition is an incarnation of the double-copy relations between open string amplitudes, Yang--Mills, and Z-theory found by Mafra, Schlotterer, and Stieberger \cite{Mafra:2011nv,Mafra:2011nw}.

In the above formula \eqref{intro1}, we have used a new CHY ingredient denoted as $\PTap(\beta)$, which we call the \emph{string Parke--Taylor} factor, since to leading order it reduces to the usual Parke--Taylor CHY integrand. It itself takes a form of a string integral with a disk ordering $\beta$,
\bes\label{intro2}
\PTap(\beta) = \int_{D(\beta)} \KN\; {\det}^\prime \Phi(\sigma, z).
\ens
Here, $D(\beta)$ denotes the disk integration region $z_{\beta(1)} < z_{\beta(2)} < \cdots < z_{\beta(n)}$ and $\KN$ is the usual Koba--Nielsen integration measure. The string Parke--Taylor factor inherits the permutation $\beta$ as a colour ordering. The key object in its definition is a determinant of an $(n-3)! \times (n-3)!$ matrix $\Phi(\sigma,z)$, which will be given in the following sections. It is similar in form to the Hodges matrix calculating MHV gravity amplitudes \cite{Hodges:2012ym}. In fact, the matrix $\Phi(\sigma,z)$ in the form used in this work has been first studied by Stieberger and Taylor in \cite{Stieberger:2013hza,Stieberger:2013nha}, who used it to derive novel relations between open superstring and supergravity amplitudes. Existence of such an object was also conjectured by Cachazo, He, and Yuan in \cite{Cachazo:2013gna}.

The goal of this note is to give a well-defined formulation of the string Parke--Taylor factor \eqref{intro2}, as well as study its properties. In particular, we show how to define an $\slc$-covariant object ${\det}^\prime \Phi(\sigma,z)$ in the case when it is not fully supported on scattering equations. We also show how different string-like models can be expressed in the CHY formalism using the string Parke--Taylor.

We organize the paper as follows. After reviewing the CHY formalism and establishing notation in section \ref{sec:ReviewOfCHY}, we give a definition of the string Parke--Taylor factor in section \ref{sec:StringParkeTaylor}. Details of the complimentary proofs are given in appendix \ref{sec:ProofOfIndependence}, and in appendix \ref{sec:BuildingBlockB} we consider abelianized version of the string Parke--Taylor. In section \ref{sec:OpenString} we show how the new ingredient can be used in CHY formulae to obtain tree-level S-matrices of open string theory and related models. We conclude in section \ref{sec:OpenQuestions} with discussion and open questions.

\section{\label{sec:ReviewOfCHY}Review of the CHY Formalism and KLT Relations}

The CHY formulation of scattering amplitudes \cite{Cachazo:2013gna,Cachazo:2013hca,Cachazo:2013iea,Cachazo:2014nsa,Cachazo:2014xea} utilizes a Riemann surface as an auxiliary object encoding the singularity structure of the S-matrix. The connection is given by the so-called \emph{scattering equations}, valid in any number of space-time dimensions,
\bes
E_a = \sum_{b \neq a} \frac{s_{ab}}{\sigma_a - \sigma_b} = 0, \qquad a=1,2,\dots,n.
\ens
Here, $k_a$ are ingoing massless momenta with $s_{ab} = (k_a + k_b)^2$, and $\sigma_a$ denote the positions of punctures on a Riemann sphere associated to the particle $a$. It is known \cite{Cachazo:2013gna} that only $n-3$ of these equations are independent, producing exactly $(n-3)!$ solutions for $\{\sigma\}$. Extensions to massive theories \cite{Dolan:2013isa,Naculich:2014naa,Naculich:2015zha}, loop level \cite{Geyer:2015bja,Geyer:2015jch,Geyer:2016wjx,Cardona:2016bpi,Cardona:2016wcr,Gomez:2016cqb}, and specific dimensions \cite{Cachazo:2013iaa,Geyer:2014fka,He:2016vfi,Cachazo:2016ror} have been considered, but for the purpose of this work, we will focus on the simplest case of massless scattering at tree level.

Scattering equations are independent of the theory under consideration. The connection to a specific theory is given by the CHY integral \cite{Cachazo:2013hca,Cachazo:2013iea}. The prescription for calculating an $n$-point amplitude is to integrate over the moduli space of genus-zero Riemann surfaces with $n$ punctures on the support of the scattering equations,
\bes\label{IntegralForm}
{\cal{A}}_n^{\rm theory} = \int \frac{d^n \sigma}{\vol\slc}\, {\prod_{a}}^\prime \delta( E_a )\; {\cal I}^{\rm theory}\! \left( \{ \sigma,k,\epsilon \} \right).
\ens
Here, we mod out the $\slc$ redundancy by fixing three punctures, say $i,j,k$. Similarly, we remove three redundant scattering equations, say $r,s,t$, which is denoted by a prime. As a consequence of these procedures, we pick up a Faddeev--Popov factor $(\sigma_{ij} \sigma_{jk} \sigma_{ki}) (\sigma_{rs} \sigma_{st} \sigma_{tr})$, where $\sigma_{ab} = \sigma_{a} - \sigma_{b}$.

The crucial requirement is that the integrand ${\cal I}^{\rm theory} \left( \{ \sigma,k,\epsilon \} \right)$ is a \emph{local} object. The entire information about the singularities of the amplitude comes from the scattering equations.

Despite its name, there are no integrations to be done in the CHY integral. It fully localizes on the solutions to the scattering equations:
\bes\label{SumOverSolutions}
{\cal{A}}_n^{\rm theory} = \sum_{m=1}^{(n-3)!} \frac{{\cal I}^{\rm theory}\! \left( \{ \sigma,k,\epsilon \} \right)}{{\det}^\prime \Phi}  \Bigg|_{\sigma_a = \sigma_a^{(m)}},
\ens
where $\sigma_{a}^{(m)}$ denotes the $m$-th solution for puncture $a$, and we have picked up a Jacobian factor, defined as
\be\label{CHYmeasure}
{\det}^\prime \Phi = \frac{(-1)^{i+j+k+r+s+t}}{(\sigma_{ij} \sigma_{jk} \sigma_{ki}) (\sigma_{rs} \sigma_{st} \sigma_{tr})} |\Phi|^{ijk}_{rst}.
\en
We have indicated that columns $i,j,k$ and rows $r,s,t$ need to be removed from the matrix before taking the determinant. It was shown \cite{Cachazo:2012da} that the above definition is independent of the the choice of $i,j,k,r,s,t$. In explicit form, the matrix reads
\bes\label{Jacobian}
\Phi_{ab} = \frac{\partial E_a}{\partial \sigma_b} =
\begin{dcases}
	\frac{s_{ab}}{(\sigma_{a} - \sigma_{b})^2} &\quad {\rm if}\quad a \neq b, \\
	-\sum_{c \neq a}\frac{s_{ac}}{(\sigma_{a} - \sigma_{c})^2} &\quad {\rm if}\quad a = b.
\end{dcases}
\ens
Despite the usefulness of the form \eqref{SumOverSolutions} in practical calculations, for the purposes of manipulating the object it is usually easier to stick with the integral form \eqref{IntegralForm}.

For all quantum field theories considered so far, the CHY integrand factors into two parts, ${\cal I}^{\rm theory} = {\cal I}^{\rm theory}_L {\cal I}^{\rm theory}_R $. In order for the integral to be $\slc$-invariant, each of the half-integrands needs to transform with weight $2$ under an $\slc$ transformation. The simplest function with this property one can write down is the \emph{Parke-Taylor} factor,
\bes
\PT(\I_n) = \frac{1}{\sigma_{12}\, \sigma_{23}\, \cdots\, \sigma_{n1}}.
\ens
It borrows its name from the famous MHV gluon amplitude found by Parke and Taylor \cite{Parke:1986gb}, which has a similar form. Here, we have opted for a canonical permutation $\I_n = (1,2,\ldots,n)$, but an object with an arbitrary permutation $\PT(\alpha)$ is defined analogously. The simplest example utilizing the Parke--Taylor factor is
\bes\label{Biadjoint}
m(\alpha | \beta) = \int\CHY\, \PT(\alpha)\, \PT(\beta),
\ens
where we have defined the CHY measure as
\bes
\CHY = \left(\sigma_{ij} \sigma_{jk} \sigma_{ki}\!\! \prod_{a \neq i,j,k}\!\!\! d \sigma_a \right)\left(\sigma_{rs} \sigma_{st} \sigma_{tr}\!\!\! \prod_{a \neq r,s,t}\!\!\! \delta(E_a) \right).
\ens
The object $m(\alpha | \beta)$ turns out to compute partial amplitudes of the bi-adjoint scalar theory \cite{Cachazo:2013iea,BjerrumBohr:2012mg}, equal to a sum over all cubic Feynman diagrams consistent with two planar orderings, $\alpha$ and $\beta$.

There is a long list of other quantum field theories one can describe within the CHY formalism \cite{Cachazo:2013iea,Cachazo:2014nsa,Cachazo:2014xea,Cachazo:2016njl,He:2016vfi,He:2016dol,Cachazo:2013iaa,He:2016iqi}, most notably Einstein gravity, Yang--Mills theory, and the non-linear sigma model. The integrands for a subset of theories are summarized below:
\begin{center}
\begin{tabular}[!h]{c|c|c}
theory & \phantom{---}$\mathcal{I}_L$\phantom{---} & \phantom{---}$\mathcal{I}_R$\phantom{---} \tr
\hline
bi-adjoint scalar & $\PT$ & $\PT$ \tr
Yang-Mills theory & $\PT$ & $\Pfp \Psi$ \tr
Einstein gravity & $\Pfp \Psi$ & $\Pfp \Psi$ \tr
non-linear sigma model & $\PT$ & $(\Pfp \mathsf{A})^2$ \tr
special Galileon & $(\Pfp \mathsf{A})^2$ & $(\Pfp \mathsf{A})^2$
\end{tabular}
\end{center}
The precise definition of $\Pfp \Psi$ and $\Pfp \mathsf{A}$ is not important for the purpose of this work, other than noting that $\Psi$ is a matrix encoding the entire dependence on the polarization vectors for the Yang--Mills and gravity amplitudes.\footnote{For a more comprehensive list of theories and their CHY integrands, see, e.g., appendix A of \cite{Cachazo:2016njl}.} A proposal for more general amplitudes that mix different species of particles has been made in \cite{Cachazo:2016njl}.

Similarities between CHY integrands for different theories are in fact a consequence of the Kawai--Lewellen--Tye (KLT) relations \cite{Kawai:1985xq,BjerrumBohr:2010hn}. CHY formalism enables a new way of understanding them using a simple linear algebra argument. For this purpose, it is convenient to re-write the form \eqref{SumOverSolutions} as a contraction of a vector, a diagonal matrix, and another vector:
\bes
{\cal{A}}_n^{\rm theory} = \mathsf{L}_{i} \, \mathsf{\Lambda}_{ij} \, \mathsf{R}_{j},
\ens
where $i,j = 1,2,\ldots,(n-3)!$ label the solutions to the scattering equations and
\bes
\mathsf{L}_{i} = {\cal I}_L \Big|_{\sigma_a = \sigma_a^{(i)}}, \quad \mathsf{\Lambda}_{ij} = \frac{\delta_{ij}}{{\det}^\prime \Phi}\Big|_{\sigma_a = \sigma_a^{(i)}}, \quad \mathsf{R}_{j} = {\cal I}_R \Big|_{\sigma_a = \sigma_a^{(j)}}.
\ens
We can consider the classic example of KLT relations between gravity and Yang--Mills amplitudes. Let us organize the partial Yang--Mills amplitudes into a vector labelled by a set of $(n-3)!$ permutations $\alpha$. We have
\bes
{\cal A}^{\rm YM}(\alpha) = \PT_{\alpha i}\, \mathsf{\Lambda}_{ij}\, \mathsf{Pf}_{j}, \qquad {\cal M}^{\rm GR} = \mathsf{Pf}_{i}\, \mathsf{\Lambda}_{ij}\, \mathsf{Pf}_{j},
\ens
where $\PT_{\alpha i}$ and $\mathsf{Pf}_{j}$ are a matrix and a vector respectively, with obvious definitions. Gravity amplitude can be re-written as
\bes\label{KLTDerivation}
{\cal M}^{\rm GR} &=& \mathsf{Pf}_{i}\, \mathsf{\Lambda}_{ij}\, \mathsf{Pf}_{j} \tr
&=& \left( \mathsf{Pf}_{i}\, \mathsf{\Lambda}_{ij}\, \mathsf{PT}_{j \alpha}\right) \left( \mathsf{PT}_{\beta k}\, \mathsf{\Lambda}_{kl}\, \mathsf{PT}_{l \alpha}\right)^{-1} \left( \mathsf{PT}_{\beta m}\, \mathsf{\Lambda}_{mp}\, \mathsf{Pf}_{p}\right) \tr
&=& {\cal A}^{\rm YM}(\alpha)\cdot m^{-1}(\alpha | \beta) \cdot {\cal A}^{\rm YM}(\beta),
\ens
which is indeed the KLT relation. We have identified the KLT kernel as an inverse matrix of partial amplitudes of the bi-adjoint amplitudes\footnote{Since we choose the interpretation of the KLT kernel as an inverse matrix of bi-adjoint amplitudes to be more fundamental, throughout this work we use the notation $m^{-1}(\alpha | \beta)$ instead of the more conventional $S[\alpha|\beta]$.}, as in \eqref{Biadjoint}. The sum over $(n-3)!$ permutaions $\alpha$ and $\beta$ is left implicit.

The above argument can be repeated for other theories in their CHY representation, leading to a web of new double-copy relations \cite{Cachazo:2014xea,Cachazo:2016njl}. In fact, one can even replace the KLT kernel with a different object, as long as it forms an invertible $(n-3)! \times (n-3)!$ matrix. We will return to this point in section \ref{sec:OpenString}.

Finally, let us briefly review a related result known as the \emph{KLT orthogonality}, originally proposed in \cite{Cachazo:2012da} and proven in \cite{Cachazo:2013hca}. We can consider a modification of \eqref{Jacobian},
\bes\label{PhiSigmaSigmaPrime}
\Phi_{ab}(\sigma,\sigma') =  
\begin{dcases}
	\frac{s_{ab}}{(\sigma_{a} - \sigma_{b})(\sigma'_{a} - \sigma'_{b})} &\quad {\rm if}\quad a \neq b, \\
	-\sum_{c \neq a}\frac{s_{ac}}{(\sigma_{a} - \sigma_{c})(\sigma'_{a} - \sigma'_{c})} &\quad {\rm if}\quad a = b.
\end{dcases}
\ens
Here, $\{\sigma\}$ and $\{\sigma'\}$ are different solutions of the scattering equations. The matrix has four null vectors,
\bes\label{NullVectors}
v_1 = (1, 1, \ldots, 1)^\intercal, &\quad& v_2 = (\sigma_2, \sigma_3, \ldots, \sigma_{n-2})^\intercal, \tr
\quad v_3 = (\sigma'_2, \sigma'_3, \ldots, \sigma'_{n-2})^\intercal, &\quad& v_4 = (\sigma_2 \sigma'_2, \sigma_3 \sigma'_3, \ldots, \sigma_{n-2} \sigma'_{n-2})^\intercal,
\ens
where for simplicity of notation we have taken columns and rows $1,n-1,n$ to be the ones removed. As a consequence of \eqref{NullVectors}, the matrix $\Phi(\sigma,\sigma')$ has rank $n-4$, or co-rank $1$, so its determinant vanishes. However, when $\sigma = \sigma'$, the two null vectors $v_2$ and $v_3$ are the same, making the determinant of $\Phi(\sigma, \sigma)$ non-zero.

In fact, it was shown in \cite{Cachazo:2013hca} that on the support of scattering equations,
\bes
{\det}^\prime \Phi(\sigma, \sigma') = \PT^{(\sigma)}(\alpha) \cdot m^{-1} (\alpha|\beta) \cdot \PT^{(\sigma')}(\beta),
\ens
where the superscripts of the Parke--Taylor factors denote the variables it is built out of. We can think of the expression as an inner product of two Parke--Taylor factors weighted by the KLT kernel. This inner product is non-vanishing only if the two sets of punctures $\{\sigma\}$ and $\{\sigma'\}$ are the same solution of the scattering equations. This is the statement of the KLT orthogonality. We will study a further generalization of the matrix \eqref{PhiSigmaSigmaPrime} in the following section. 

\section{\label{sec:StringParkeTaylor}String Parke--Taylor Factor}

In this section we introduce a new object, called the string Parke--Taylor factor, which will enter the CHY formula for open string amplitudes as a half-integrand. Before moving on to the most general definition, we first study the simplest case in a gauge-fixed form.

\subsection{\label{PreliminaryDefinition}Preliminary Definition}

We start by defining a matrix similar to \eqref{PhiSigmaSigmaPrime},
\bes\label{PhiPreliminary}
\Phi_{ab}(\sigma,z) =  
\begin{dcases}
	\frac{\alpha' s_{ab}}{\sigma_{ab}\, z_{ab}} &\quad {\rm if}\quad a \neq b, \\
	-\sum_{c \neq a}\frac{\alpha' s_{ac}}{\sigma_{ac}\, z_{ac}} &\quad {\rm if}\quad a = b.
\end{dcases}
\ens
However, now we do not assume that $\{\sigma\}$ or $\{z\}$ are solutions of the scattering equations or satisfy any other relation. We still impose momentum conservation, $\sum_a k_a^\mu =0$, and denote $s_{ab} = (k_a + k_b)^2$. We wish to compute the object
\be\label{DeterminantPhi}
{\det}^\prime \Phi(\sigma,z) = \frac{|\Phi(\sigma,z)|^{1,n-1,n}_{1,n-1,n}}{(\sigma_{1,n-1} \sigma_{n-1,n} \sigma_{n,1}) (z_{1,n-1} z_{n-1,n} z_{n,1})},
\en
in the gauge $(\sigma_1, \sigma_n, \sigma_{n-1}) = (z_1, z_{n-1}, z_n) = (0,1,\infty)$. We will make use of the following identity:
\bes\label{DetPhiExpansion}
{\det}^\prime \Phi(\sigma,z) = (-\alpha')^{n-3} \sum_{\alpha \in S_{n-3}} \frac{1}{\sigma_{1,\alpha(2)} \sigma_{\alpha(2),\alpha(3)} \cdots \sigma_{\alpha(n-4),\alpha(n-3)}} \prod_{a=2}^{n-2} \sum_{b=1}^{a-1} \frac{s_{\alpha(b), \alpha(a)}}{z_{\alpha(b), \alpha(a)}},
\ens
where the first sum runs over the permutations of labels $(2,3,\ldots,n-2)$. Proof of this identity has been given in \cite{Stieberger:2013hza} and \cite{Cachazo:2013gna} on the support of scattering equations. However, this assumption can be easily relaxed to show that \eqref{DetPhiExpansion} holds as an algebraic identity. The proof is structurally identical to the expansion of the Hodges determinant for MHV gravity amplitudes \cite{Hodges:2012ym} given in \cite{Feng:2012sy}, and therefore skipped here.

Equipped with this knowledge, let us turn to the open string amplitude. We will make use of the celebrated formula discovered by Mafra, Schlotterer, and Stieberger \cite{Mafra:2011nv,Mafra:2011nw} using the pure spinor formalism \cite{Berkovits:2000fe},
\be\label{OpenFSYM}
{\cal A}^{\rm{open}}(\beta) = \sum_{\alpha \in S_{n-3}} F_\beta^\alpha \cdot {\cal A}^{\rm SYM}(\alpha).
\en
Here, the open superstring amplitude decomposes into two parts. The functions $F_{\beta}^{\alpha}$ carry the entire dependence on $\alpha'$, while ${\cal A}^{\rm SYM}$ carries the whole information about polarizations of the strings.\footnote{In order to keep the discussion valid in general space-time dimension, we will specialize to purely gluonic external states, and hence rename ${\cal A}^{\rm SYM} \to {\cal A}^{\rm YM}$.} It is natural to write the former as a string integral, and the latter as a CHY integral as follows:
\be
{\cal A}^{\rm{open}}(\beta) = (-\alpha')^{n-3}\!\!\!\! \sum_{\alpha \in S_{n-3}}\!\! \int_{D(\beta)} \!\!\!\!\! \KN \prod_{a=2}^{n-2} \sum_{b=1}^{a-1} \frac{s_{\alpha(b), \alpha(a)}}{z_{\alpha(b), \alpha(a)}} \!\int\! \CHY \frac{1}{\sigma_{1,\alpha(2)} \sigma_{\alpha(2),\alpha(3)} \cdots \sigma_{\alpha(n-4),\alpha(n-3)}} \Pfp \Psi, \nonumber
\en
where $D(\beta)$ denotes the integration region $0 \leq z_{\beta(2)} \leq z_{\beta(3)} \leq \cdots \leq z_{\beta(n-2)} \leq 1$, and we have defined the Koba--Nielsen measure for string integrals,
\bes\label{KNMeasure}
\KN = z_{1,n-1} z_{n-1,n} z_{n,1}\!\!\!\! \prod_{a \neq, 1,n-1,n}\!\!\!\! dz_a\; \prod_{a<b} |z_{ab}|^{\alpha' s_{ab}}.
\ens
We have also used the same gauges as above, and absorbed the Faddeev--Popov factors from the measures. One can immediately recognize the expression \eqref{DetPhiExpansion} appearing here from the two separate integrals. Substituting \eqref{DetPhiExpansion}, we obtain
\bes
{\cal A}^{\rm{open}}(\beta) = \int_{D(\beta)}\!\!\!\!\KN\, \left( \int  \CHY\, {\det}^\prime \Phi(\sigma,z)\, \Pfp \Psi \right),
\ens
which is the expression originally proposed by Cachazo, He, and Yuan in \cite{Cachazo:2013gna}. From this perspective, the CHY integral in the brackets calculates the correlation function of vertex operators needed for the open string computation modulo the Koba--Nielsen factor. Commuting the integrals gives an alternative interpretation:
\bes\label{OpenStringSecond}
{\cal A}^{\rm{open}}(\beta) =  \int \CHY \left( \int_{D(\beta)}\!\!\!\!\KN\, {\det}^\prime \Phi(\sigma,z)\right) \Pfp \Psi.
\ens
The term in the brackets now becomes a CHY half-integrand on its own right. Since the left hand side of the equation becomes a Yang--Mills amplitude in the $\alpha' \to 0$ limit, the term in the bracket necessarily becomes a Parke--Taylor factor, $\PT(\beta)$, up to terms vanishing on the support of scattering equations. The ordering is inherited from the disk integral. We conclude that the newly-found half-integrand is an $\alpha'$-corrected Parke--Taylor factor, which we denote as $\PTap(\beta)$.

So far we have been working in a specific gauge. The determinant \eqref{DeterminantPhi} is also not manifestly $\slr$- and $\slc$-covariant. We address these issues in the following subsection.

\subsection{M\"obius-invariant Definition}
\label{invariant}
In order to make the expression \eqref{PhiPreliminary} $\slr$-covariant with respect to $z$, and $\slc$-covariant with respect to $\sigma$, we introduce two reference punctures, $\sigma_q$ and $z_q$, in the diagonal terms as follows:
\bes
\Phi_{ab}(\sigma,z) :=  
\begin{dcases}
	\frac{\alpha' s_{ab}}{\sigma_{ab}\, z_{ab}} &\quad {\rm if}\quad a \neq b, \\
	-\sum_{c \neq a} \frac{\alpha' s_{ac}}{\sigma_{ac}\, z_{ac}} \frac{\sigma_{cq}}{\sigma_{aq}} \frac{z_{cq}}{z_{aq}} &\quad {\rm if}\quad a = b.
\end{dcases}
\ens
With this modification, the determinant of $\Phi(\sigma,z)$ manifestly transforms with weight $2$ in both sets of coordinates. What remains is to show that the entire object is independent of the choice of columns and rows deleted from the matrix. 
In appendix \ref{sec:ProofOfIndependence} we prove that the combination
\be\label{reddet}
{\rm det}'\Phi(\sigma,z) := \frac{(-1)^{i+j+r+s}}{\sigma_{rs}\sigma_{ik}\sigma_{jk}\, z_{ij} z_{rk}z_{sk}}|\Phi(\sigma,z)|^{ijk}_{rsk}\;\, \underset{\tiny\begin{subarray}{c}\text{e.g.}\\ r=i\\ s=j \end{subarray}}{=}\;\, \frac{|\Phi(\sigma,z)|^{ijk}_{ijk}}{(\sigma_{ij}\sigma_{jk}\sigma_{ki}) (z_{ij} z_{jk} z_{ki})}
\en
is independent of the choice of $i,j,k,r,s$, as well as the reference punctures $z_q, \sigma_q$. Since in the gauges considered in the previous subsection this expression collapses to \eqref{DeterminantPhi}, we take it as a definition of ${\rm det}'\Phi(\sigma,z)$. The reader might be tempted to consider a combination $|\Phi(\sigma, z)|^{ijk}_{rst}/(\sigma_{ij}\sigma_{jk}\sigma_{ki}z_{rs}z_{st}z_{tr})$ with all six labels distinct, in analogy with the factor appearing in the CHY measure \eqref{CHYmeasure}. Such combination, however, would not have the correct transformation properties.

We can now make a proper definition for the string Parke--Taylor factor:
\begin{empheq}[box=\fbox]{align}\label{StringPTDefinition}
\;\PTap(\beta) := \int_{D(\beta)} \KN\; {\det}^\prime \Phi(\sigma, z).\;
\end{empheq}
Let us briefly remark on convergence of this integral. In \cite{Cachazo:2016ror}, together with Cachazo we have studied a region of the space of kinematic invariants, called ${\cal K}^+_n$, defined by $s_{ij} > 0$ for $1 \leq i < j \leq n-1$ but $(i,j) \neq (1,n-1)$. We have shown that such kinematics can be realized in complexified four-dimensional space-time. In conjunction with the gauge fixing of $z_1, z_{n-1}, z_n$, this region makes the integrals \eqref{StringPTDefinition} convergent for any permutation $\beta$. As it is conventional in string theory \cite{Schwarz:1982jn}, we define \eqref{StringPTDefinition} in other kinematic regions as an analytic continuation of the result in ${\cal K}^+_n$. Interestingly, in the same kinematic region, for each domain of integration $D(\beta)$, the string Parke--Taylor has exactly one saddle point when $\alpha' \to \infty$ coinciding with one solution of the scattering equations \cite{Cachazo:2016ror}. On these saddle points \eqref{reddet} cancels out with the CHY Jacobian \eqref{CHYmeasure}, giving the correct $\alpha' \to \infty$ limit \cite{Fairlie:1972zz,RobertsPhD,Gross:1987ar}.

\subsection{Properties and Examples}

We can now turn to studying properties of the string Parke--Taylor factor \eqref{StringPTDefinition}. Since its integrand is permutation invariant, the following properties follow purely as a consequence of the disk integration $D(\beta)$,
\bes\label{Properties}
{\rm Cyclicity:} \quad && \PTap(1,2,3,\ldots,n) = \PTap(2,3\ldots,n,1),\\
{\rm Reflection:} \quad && \PTap(1,2,\ldots,n) = (-1)^n\, \PTap(n,\ldots,2,1), \phantom{\sum^{n}}\\
{\rm Monodromy:} \quad && \sum_{i=2}^{n} e^{i \pi \alpha' k_1 \cdot (k_2 + \ldots + k_i)} \,\PTap(2,3,\ldots,i,1,i+1,\ldots,n) = 0.\qquad
\ens
In Subsection \ref{PreliminaryDefinition}, we have argued that $\PTap$ reduces to the field-theory Parke--Taylor in the $\alpha' \to 0$ limit. The next question is how to expand this object to higher orders in $\alpha'$. Fortunately, we can use \eqref{OpenFSYM} and \eqref{OpenStringSecond} in order to rewrite it in terms of the familiar matrix $F_\beta^\alpha$,
\bes\label{PTAlphaPrimeFPT}
\PTap(\beta) = \sum_{\alpha \in S_{n-3}} F_\beta^\alpha \cdot \PT(\alpha)\;\; = \sum_{\alpha, \gamma \in S_{n-3}}\!\! Z_\beta(\alpha) \cdot m^{-1}(\alpha | \gamma) \cdot \PT(\gamma).
\ens
In the second line we have expanded $F_\beta^\alpha$ in terms of the so-called Z-theory amplitudes \cite{Carrasco:2016ldy,Mafra:2016mcc,Carrasco:2016ygv}. We can now use known results about the expansion of $F_\beta^\alpha$ using Drinfeld associator \cite{Broedel:2013aza}, or expansion of $Z_\beta(\alpha)$ with the Berends--Giele method of \cite{Mafra:2016mcc}. For $n=3,4,5$ we have the exact results:
\bes
\PTap(\I_3) &=& \PT(\I_3),\\
\label{PT4}\PTap(\I_4) &=& \PT(\I_4) \frac{\Gamma(1+\alpha' s) \Gamma(1+ \alpha' t)}{\Gamma(1-\alpha' u)} \tr
&=& \PT(\I_4) \left( 1 - \alpha'^2 \zeta_2\, s t - \alpha'^3 \zeta_3\, s t u - \alpha'^4 \zeta_4\, s t (s^2 + st/4 + t^2) +  \dots \right),\qquad\qquad\qquad\quad
\ens
where $s = s_{12}$, $t = s_{23}$, $u = s_{13}$. Setting $\alpha'=1$ for clarity we also have
\bes\label{PT5}
\PTap(\I_5) &=& \PT(\I_5)\, \Gamma(1+s_{12}) \Gamma(1+s_{23}) \Gamma(1+s_{34}) \Gamma(1+s_{45})\tr
&&\times \Bigg( \phantom{}_3 \tilde{F}_2 \left(s_{12}, 1+s_{45},-s_{24}; 1+s_{12} + s_{23}, 1+s_{34}+s_{45} ;1\right)\tr
&&\qquad- s_{13}s_{24}\, \frac{\sigma_{12} \sigma_{34}}{\sigma_{13} \sigma_{24}}\, \phantom{}_3 \tilde{F}_2 \left(1+s_{12}, 1+s_{45},1-s_{24}; 2+s_{12} + s_{23}, 2+s_{34}+s_{45} ;1\right)  \Bigg)\tr
&=& \PT(\I_5) \Bigg( 1 - \alpha'^2 \zeta_2 \Bigg( -s_{12}s_{34} + s_{34}s_{45} + s_{12} s_{51} + s_{13}s_{24}\frac{\sigma_{12} \sigma_{34}}{\sigma_{13} \sigma_{24}} \Bigg) \tr
&&\qquad\qquad+ \alpha'^3 \zeta_3 \Bigg( s_{12} \left(-s_{34}^2 - s_{12} s_{34} - 2 s_{23} s_{34} + s_{12}s_{51} + s_{51}^2\right) +s_{34} s_{45}(s_{34}+s_{45}) \tr
&&\qquad\qquad\qquad\qquad  + s_{13}s_{24} (s_{12} + s_{23} + s_{34} + s_{45} + s_{51}) \frac{\sigma_{12} \sigma_{34}}{\sigma_{13} \sigma_{24}} \Bigg) + \ldots \Bigg),
\ens
where $\phantom{}_3 \tilde{F}_2(a,b,c;d,e;z) = \phantom{}_3 F_2(a,b,c;d,e;z)/\Gamma(d)\Gamma(e)$ is a regularized hypergeometric function. Expanding the $6$-pt case we obtain:
\bes\label{PT6}
\PTap(\I_6) &=& \PT(\I_6) \Bigg( 1 - \alpha'^2 \zeta_2 \Bigg( s_{12} s_{61} - s_{45}(s_{45}+s_{46}) + s_{345}(s_{13}+s_{23})\tr
&&\qquad\qquad\qquad\qquad\quad + s_{14}s_{25} \frac{\sigma_{12}\sigma_{45}}{\sigma_{14}\sigma_{25}} + s_{13}(s_{24}+s_{25}) \frac{\sigma_{12}\sigma_{34}}{\sigma_{13}\sigma_{24}} + s_{35}(s_{14}+s_{24}) \frac{\sigma_{23} \sigma_{45}}{\sigma_{24}\sigma_{35}}\tr
&&\qquad\qquad\qquad\qquad\quad + s_{13}s_{25} \frac{\sigma_{12}\sigma_{23}\sigma_{45}}{\sigma_{13}\sigma_{24}\sigma_{25}} + s_{14}s_{35} \frac{\sigma_{12}\sigma_{34}\sigma_{45}}{\sigma_{14}\sigma_{24}\sigma_{35}}\Bigg) + \ldots \Bigg).
\ens
As confirmed by these examples, $\PTap$ has a local form as an expansion in $\alpha'$.

Of course, it would be beneficial to find closed-form formulae for the $\alpha'$-expansion for arbitrary $n$. For example, since the ${\cal O}(\alpha'^2)$ correction to the open superstring comes from the ${\rm Tr} F^4$ operator, we can use the result of He and Zhang \cite{He:2016iqi} to write the expansion to the sub-leading order:
\bes\label{HeZhang}
\PTap(\I_n) = \PT(\I_n) \left(1 + \alpha'^2 \zeta_2\!\!\!\! \sum_{i<j<k<l}\!\!\!\! s_{jk} s_{li}\, \frac{\sigma_{ij} \sigma_{kl}}{\sigma_{jk}\sigma_{li}} + \ldots \right).
\ens
This expression can be shown to be equivalent to (\ref{PT4}--\ref{PT6}) using cross-ratio identities \cite{Cardona:2016gon}. The existence of such a closed-form formula suggests that higher-order expansion could take a similarly succinct form.

Since both soft and factorization limits of open superstring match those of its field-theory equivalent \cite{Mafra:2011nw}, the string Parke--Taylor factor has to have the same behaviour as the field-theory Parke--Taylor in these limits.

\section{\label{sec:OpenString}Open String Amplitudes in the CHY Formalism}

In the previous section we have shown that the open superstring amplitudes in the gluon sector can be written in the CHY language as
\begin{empheq}[box=\fbox]{align}\label{OpenCHY}
\;{\cal A}^{\rm open}(\beta) = \int \CHY\; \PTap(\beta)\; \Pfp \Psi.\;
\end{empheq}
This structure is reminiscent of \eqref{OpenFSYM}, since $\PTap$ carries all the $\alpha'$-dependence and information about the disk, and $\Pfp \Psi$ knows about the polarization degrees of freedom. We would like to find out to what extent this construction generalizes to other string-like integrals studied in the literature \cite{Carrasco:2016ldy,Mafra:2016mcc,Mizera:2016jhj,Carrasco:2016ygv,Huang:2016tag}.

In this section we will make use of one of the major strengths of the CHY formalism, which is the way it implements KLT relations. Indeed, the linear algebra argument behind the original KLT relation between gravity and Yang--Mills \eqref{KLTDerivation} can be generalized to other cases. We can write it schematically as
\bes\label{CHYSchematic}
\int \CHY\, {\cal I}_L\, {\cal I}_R = \left( \int \CHY\, {\cal I}_L\, {\cal I}_A \right)\cdot \left( \int \CHY\, {\cal I}_B\, {\cal I}_A \right)^{\!\!-1} \!\!\!\!\!\cdot \left( \int \CHY\, {\cal I}_B\, {\cal I}_R \right).
\ens 
Here, the only requirement is that ${\cal I}_A$ and ${\cal I}_B$ have some additional label that allows for a construction of an invertible $(n-3)! \times (n-3)!$ matrix out of its amplitudes. In the case of the field-theory KLT, these labels are the permutations of the Parke--Taylor half-integrand, i.e., ${\cal I}_A = \PT(\alpha)$ and ${\cal I}_B = \PT(\beta)$ for $\alpha$ and $\beta$ belonging to some sets of $(n-3)!$ permutations.

One immediate application of this fact is finding a representation of the Z-theory amplitudes \cite{Mafra:2016mcc,Carrasco:2016ldy,Carrasco:2016ygv}. In order to satisfy the KLT relation ${\cal A}^{\rm open}(\beta) = Z_{\beta} (\gamma) \cdot m^{-1}(\gamma | \delta) \cdot {\cal A}^{\rm YM}(\delta)$, it must be that
\begin{empheq}[box=\fbox]{align}\label{ZCHY}
\;Z_{\beta}(\gamma) = \int \CHY\; \PTap(\beta)\; \PT(\gamma).\;
\end{empheq}
One can also obtain abelianized Z-theory amplitudes \cite{Carrasco:2016ldy,Carrasco:2016ygv}, by using an abelianized string Parke--Taylor factor, $\PTap(\times_n)$, whose definition is given in appendix \ref{sec:BuildingBlockB}. Repeating the same procedure for the string amplitude \eqref{OpenCHY} yields the full string theory version of the Born--Infeld theory.

Let us turn to studying the string theory KLT kernel \cite{Kawai:1985xq,BjerrumBohr:2010hn}. In \cite{Mizera:2016jhj} one of the authors argued that inverse of this kernel, denoted by $m_{\alpha'}(\beta | \tilde{\beta})$, reveals a surprisingly simple structure, akin to the one of the field-theory bi-adjoint scalar. It was also conjectured that such an object comes from an auxiliary string-like integral with two disk orderings $\beta$ and $\tilde{\beta}$, related to Z-theory by a KLT relation $m_{\alpha'}(\beta | \tilde{\beta}) = Z_{\beta}(\gamma) \cdot m^{-1}(\gamma | \tilde{\gamma}) \cdot \overline{Z_{\tilde{\beta}}(\tilde{\gamma})}$. The overbar denotes a chiral deformation \cite{Huang:2016bdd,Leite:2016fno}, equivalent to taking $\alpha' \to -\alpha'$. Using this relationship, we identify
\begin{empheq}[box=\fbox]{align}\label{MAlphaPrime}
\;m_{\alpha'}(\beta | \tilde{\beta}) = \int \CHY\; \PTap(\beta)\; \PT_{-\alpha'}(\tilde{\beta}).\;
\end{empheq}
In this case, second string Parke--Taylor factor is taken to have a negative string tension. The object \eqref{MAlphaPrime} is symmetric with respect to both permutations. As was checked explicitly in \cite{Mizera:2016jhj}, all odd transcendentality cancels between the two string Parke--Taylor factors, and the final answer can be written in a compact form built purely out of trigonometric functions. It is straightforward to rewrite \eqref{MAlphaPrime} as a string integral with two disks.

The new CHY construction gives a way of incorporating the results of \cite{Huang:2016bdd}, where the full string theory KLT of an open superstring with a chiral open superstring leads to a cancellation of all massive poles, giving pure gravity. In fact, the form of \eqref{OpenCHY}, \eqref{ZCHY}, and \eqref{MAlphaPrime} means we can write pure gravity as a KLT in at least three distinct ways:
\bes
{\cal M}^{\rm GR} &=& {\cal A}^{\rm YM}(\alpha)\cdot m^{-1}(\alpha | \beta) \cdot {\cal A}^{\rm YM}(\beta) \tr
&=& {\cal A}^{\rm open}(\alpha)\cdot m_{\alpha'}^{-1}(\alpha | \beta) \cdot \overline{{\cal A}^{\rm open}(\beta)} \tr
&=& {\cal A}^{\rm open}(\alpha)\cdot Z_{\alpha}^{-1}(\beta) \cdot {\cal A}^{\rm YM}(\beta).
\ens
So the Z-theory can also be used as a type of a KLT kernel translating between string and field-theory.

Yet another consequence of \eqref{MAlphaPrime} is the expansion identity for the string Parke--Taylor factor,
\bes
\PTap(\beta) = m_{\alpha'}(\beta | \tilde{\beta}) \cdot m_{\alpha'}^{-1}(\tilde{\beta} | \gamma) \cdot \PTap(\gamma),
\ens
which is a way of solving monodromy relations \eqref{Properties} for an $(n-3)!$ basis of string Parke--Taylors. Together with its field-theory equivalent \cite{Cachazo:2013iea}, which can be obtained in the $\alpha' \to 0$ limit, they become a convenient way of changing bases for both disk and Parke--Taylor orderings \cite{Mizera:2016jhj}. 

Finally, we briefly consider the open bosonic string amplitudes as well. The factorization similar to the superstring case, ${\cal A}^{\rm open}_{\rm bosonic}(\alpha) = Z_{\alpha}(\beta) \cdot m^{-1}(\beta | \gamma) \cdot B(\gamma)$, found in \cite{Huang:2016tag}, suggests that there should exist a corresponding CHY formula. Although we do not have the full answer, we can borrow the result of \cite{He:2016iqi} for the ${\rm Tr}F^3$ amplitudes in order to write the leading expansion:
\begin{empheq}[box=\fbox]{align}
\;{\cal A}^{\rm open}_{\rm bosonic}(\beta) = \int \CHY\; \PTap(\beta)\; (\Pfp \Psi + \alpha' {\cal P}_n + \ldots),\;
\end{empheq}
where the form of ${\cal P}_n$ is given in \cite{He:2016iqi,Zhang:2016rzb}. This also gives a natural proposal for the object $B(\alpha)$ studied in \cite{Huang:2016tag}:
\be\label{BCHY}
B(\alpha) = \int \CHY\; \PT(\alpha)\; (\Pfp \Psi_n + \alpha' {\cal P}_n + \ldots).
\en

We summarize all the findings of this section in a table giving CHY integrands for various string-like theories:
\begin{center}
	\begin{tabular}[!h]{c|c|c}
		theory & \phantom{---}$\mathcal{I}_L$\phantom{---} & \phantom{---}$\mathcal{I}_R$\phantom{---} \tr
		\hline
		open superstring & $\PTap$ & $\Pfp \Psi$ \tr
		open bosonic string & $\PTap$ & $\Pfp \Psi + \alpha' {\cal P}_n + \ldots$ \tr
		Z-theory & $\PTap$ & $\PT$ \tr
		inverse KLT kernel & $\PTap$ & $\PT_{-\alpha'}$ \tr
		string Born--Infeld & $\PTap(\times)$ & $\Pfp \Psi$ \tr
		abelianized Z-theory & $\PTap(\times)$ & $\PT$
	\end{tabular}
\end{center}
In principle, using above building blocks one could try to construct new string-like models. For example, using a combination of the abelianized string Parke--Taylor factor $\PTap(\times_n)$ and $(\Pfp {\mathsf A})^2$ would give one possibility for obtaining an $\alpha'$-completion of the special Galileon theory. It is important to mention, however, that not every $\alpha'$-deformation of an amplitude necessarily comes from a consistent string theory. 

\section{\label{sec:OpenQuestions}Discussion}

In this work we have studied properties of the string Parke--Taylor factor, defined as
\bes\label{stringPT-final}
\PTap(\beta) \;=\; \int_{D(\beta)} \!\! \KN\; {\det}^\prime \Phi(\sigma, z) \;=\; \frac{1}{\sigma_{\beta(1) \beta(2)} \sigma_{\beta(2) \beta(3)} \cdots \sigma_{\beta(n) \beta(1)}} \;+\; {\cal O}(\alpha'^2).
\ens
We showed how to define the reduced determinant ${\det}^\prime \Phi(\sigma, z)$ in an $\slc$-covariant manner, as well as how to express open string amplitudes in the CHY formalism using the string Parke--Taylor factor.

Several questions about the nature of $\PTap$ remain open. Note how to leading order in $\alpha'$, the string Parke--Taylor factor gives a correlation function of free fermions with positions given by $\{\sigma\}$. Introducing $\alpha'$ corrections can be thought of as switching on interactions between the fermions, with the Mandelstam invariants $\{s_{ab}\}$ playing the r\^ole of strength couplings. It would be very interesting to find an interpretation of the string Parke--Taylor as computing some sort of correlation functions on a Riemann sphere, in particular in the context of ambitwistor strings \cite{Mason:2013sva}.

Additionally, it would be beneficial to find other ways of consistently expanding the object $\PTap$ in $\alpha'$, perhaps following the approach similar to the one employed in \cite{Boels:2008fc}. We hope that this could pave a way to finding compact expressions for string-corrected CHY amplitudes, similar to those of He and Zhang \cite{He:2016iqi,Zhang:2016rzb}. We leave these interesting questions for future research.

Another interesting question is to investigate whether one-loop amplitudes of open strings can be expressed in a language similar to \eqref{intro1}. CHY formalism admits a one-loop analog with the underlying Riemann sphere replaced with a nodal Riemann sphere \cite{Geyer:2015bja,Geyer:2015jch}. One way or another, adding $\alpha'$ corrections would have to amount to resolving this singularity, so that the nodal sphere becomes a torus.

\section*{Acknowledgements}
We are grateful to Freddy Cachazo for constant encouragement throughout this work and insightful comments. We would like to thank P. Benincasa and O. Schlotterer for useful discussions, and in particular C. Kalousios for discussions on the computation of higher-order terms in \eqref{HeZhang}. This research was supported in part by Perimeter Institute for Theoretical Physics. Research at Perimeter Institute is supported by the Government of Canada through the Department of Innovation, Science and Economic Development Canada and by the Province of Ontario through the Ministry of Research, Innovation and Science.

\appendix

\section{\label{sec:ProofOfIndependence}Definition of ${\det}^\prime \Phi(\sigma, z)$}

In this appendix we prove that the combination
\be\label{detPhi}
{\det}^\prime \Phi(\sigma,z) = \frac{(-1)^{i+j+r+s}}{\sigma_{rs}\sigma_{ik}\sigma_{jk}\, z_{ij} z_{rk}z_{sk}}|\Phi(\sigma,z)|^{ijk}_{rsk},
\en
is independent of the choice of labels $i,j,k,r,s \in \{1,2,\ldots, n\}$, as well as of the reference punctures $\sigma_q$ and $z_q$.\footnote{Note that $\det^\prime \Phi(\sigma, z)$ as defined here appears to be antisymmetric with respect to relabelling, e.g. $i \leftrightarrow j$. This is because it should really be thought of as being multiplied by differential forms coming from CHY and Koba--Nielsen integration measures, which restores the symmetry.} See section \ref{invariant} for a definition of the matrix $\Phi(\sigma,z)$. It is important to clarify that for the proofs it is not enough to simply replace integration by parts (IBP) identities by assuming support of scattering equations in variables $\{z\}$.

We split the proof into three parts: Firstly, we prove independence of $i,j,r,s$ for fixed $k$ and the choice $\sigma_q=\sigma_k, z_q = z_k$. Secondly, we prove independence of the reference punctures $\sigma_q$ and $z_q$ for generic $i,j,k,r,s$. Finally, we show that \eqref{detPhi} is independent of the choice of the label $k$.

For the purpose of the proofs, we use the notation where $A^{[\alpha,\beta,\ldots]}_{[\gamma,\delta,\ldots]}$ denotes the matrix $A$ with rows $\alpha,\beta,\ldots$ and columns $\gamma,\delta,\ldots$ removed. We also shorten $\Phi(\sigma,z)$ to just $\Phi$ for the sake of brevity. For example, writing $\det \Phi^{[ijk]}_{[rsk]}$ is equivalent to $|\Phi(\sigma,z)|^{ijk}_{rsk}$.

\subsection{Independence of the Choice of Labels $i,j,r,s$}
Without loss of generality, let us choose the label $k$ to be $n$. We also choose special values for reference punctures, $\sigma_q=\sigma_n$ and $z_q=z_n$. Since ${\det}^\prime \Phi(\sigma, z)$ is $\slc \times \slr$ covariant, it is enough to prove independence of the choice of $i,j,r,s$ at the point where $z_n\to \infty$ and $\sigma_n \to \infty$. With this gauge, the proof simplifies to that of the object:
\be
\frac{(-1)^{i+j+r+s}}{\sigma_{rs}\, z_{ij}}|\Phi|^{ijn}_{rsn}.
\en
The matrix $\Phi^{[n]}_{[n]}$ has two null vectors: $(1,1,\dots,1)^\intercal$ and $(z_1,z_2,\dots, z_{n-1})^\intercal$, on the support of scattering equations and using momentum conservation. The factor $(-1)^{i+j}/z_{ij}$ is then simply the Fadeev--Popov determinant for these null vectors. We can therefore fix $i,j$ and concentrate on the remaining invariance of $r,s$. It amounts to showing that
\be\label{claim}
\frac{(-1)^{r+s}}{\sigma_{rs}}|\Phi|^{ijn}_{rsn} \;\ibp\; \frac{(-1)^{r'+s'}}{\sigma_{r's'}}|\Phi|^{ijn}_{r's'n},
\en
which should hold on the support of integration by parts (IBP) identities. Recall that the integration measure $d\mu^{\rm KN}$ contains a Koba--Nielsen factor, which should be taken into account when using the IBP relations. Since columns and rows can be sequentially exchanged one-by-one, it is enough to show the case $r' = r,\, s' = s+1$:
\be\label{ibp}
\sigma_{r,s+1}|\Phi|^{ijn}_{rsn} \;+\; \sigma_{rs}|\Phi|^{ijn}_{r,s+1,n} \,\ibp\, 0.
\en
Let us choose, without loss of generality, the labels $i=n-2$ and $j=n-1$ to simplify the notation. Then, we write the matrix $\Phi^{[i,j,n]}_{[r,n]}$ as a set of column vectors:
\be\label{Phi}
\Phi^{[i,j,n]}_{[r,n]} = ({\bf v}_1, {\bf v}_2, \ldots, \hat{\bf v}_r, \ldots, {\bf v}_s, {\bf v}_{s+1}, \ldots, {\bf v}_{n-1}),
\en
where $\hat{\bf v}_r$ means the $r$-th column has been removed. Using the definition of the matrix $\Phi(\sigma,z)$ from \eqref{PhiPreliminary} with $\alpha' = 1$, we find the following combinations of the column vectors:
\be
\sum_{b=1}^{n-1}{\bf v}_b = {\bf 0},\qquad \sum_{b=1}^{n-1}\sigma_b{\bf v}_b = - \Big( \sum_{c \neq 1} \frac{s_{1c}}{z_{1c}} \,,\,\dots\,, \sum_{c \neq n-1} \frac{s_{n-1,c}}{z_{n-1,c}}\Big)^\intercal = -(\partial_1 \log \kn,\dots, \partial_{n-3} \log \kn)^\intercal,\nonumber
\en
where we used the definition $\kn := \prod_{a<b}|z_{ab}|^{s_{ab}}$ and a shorthand notation $\partial_a := \partial/\partial z_a$. We can use the above column vectors to define a new one:
\be\label{basic}
{\bf T} :=\kn \sum_{b=1}^{n-1}\sigma_{rb}{\bf v}_b =\kn \left( \sigma_r \sum_{b=1}^{n-1}{\bf v}_b - \sum_{b=1}^{n-1}\sigma_b{\bf v}_b \right) = (\partial_1\kn,\dots, \partial_{n-3}\kn)^{\sf T},
\en
where we have absorbed the $\kn$ factor from the measure $\KN$ into ${\bf T}$. We define an $(n-3)\times (n-3)$ matrix $\widetilde{\Phi}$, which is \eqref{Phi} with columns ${\bf v}_s, {\bf v}_{s+1}$ replaced by $\bf T$:
\be\label{phit}
\widetilde{\Phi} := ({\bf v}_1,{\bf v}_2,\ldots,\hat{\bf v}_r,\ldots,{\bf v}_{s-1},{\bf T},{\bf v}_{s+2}, \ldots, {\bf v}_{n-1}).
\en
Proving the statement \eqref{ibp} is then equivalent to showing that $\det \widetilde{\Phi}$ integrates to zero, since the LHS of \eqref{ibp} gives:
\be
\det ({\bf v}_1, {\bf v}_2, \ldots, \hat{\bf v}_r, \ldots,\, \sigma_{rs}{\bf v}_s+\sigma_{r,s+1}{\bf v}_{s+1},\, \ldots, {\bf v}_{n-1}).
\en
Multiplying the other column vectors ${\bf v}_i$ by $\sigma_{ri}$ for $i=1,2,\ldots,\hat{r},\dots,\hat{s},\widehat{s+1},\dots,n-1$ and adding them to $\sigma_{rs}{\bf v}_s+\sigma_{r,s+1}{\bf v}_{s+1}$ yields ${\bf T}$. The exact form of $\kn$ in the vector ${\bf T}$ will not matter in our proof, as long as it vanishes on the boundary.

In the following, we will assume that punctures $z_i, z_j, z_n$ are gauge-fixed in order to perform IBP. However, since the whole expression for the string Parke--Taylor factor is $\slr$-invariant, it must hold in any gauge. Let us work out a couple of low-point examples first. For $n=4$ we have:
\be
\det \widetilde{\Phi} = \partial_1 \kn,
\en
which is a total derivative with respect to $z_1$ and hence vanishes. For the case of $n=5$ with $r=2, s=3$ we obtain:
\be\label{ex-5pt}
\setlength{\arraycolsep}{5pt}
\det \widetilde{\Phi} = \begin{vmatrix}
	\Phi_{11} & \partial_1 \kn \\
	\Phi_{12} & \partial_2 \kn \\
\end{vmatrix} = \begin{vmatrix}
\Phi_{11}^{\mathsf{R}} + \Phi_{11}^{\mathsf{I}} & \partial_1 \kn \\
\Phi_{12} & \partial_2 \kn \\
\end{vmatrix},
\en
where we have split the entry $\Phi_{11}$ into two parts: \emph{relevant} $\Phi_{11}^{\mathsf{R}}$, which is independent of $z_i,z_j,z_n$, and \emph{irrelevant} $\Phi_{11}^{\mathsf{I}}$, which contains the remaining contributions depending on $z_i,z_j$ and $z_1$. Explicitly we have: $\Phi_{11}^{\mathsf{R}} = -s_{12}/(\sigma_{12} z_{12}) = -\Phi_{12}$ and $\Phi_{11}^{\mathsf{I}} = -s_{13}/(\sigma_{13} z_{13}) - s_{14}/(\sigma_{14} z_{14}) = -\Phi_{13} - \Phi_{14}$. The irrelevant term gives a vanishing contribution to \eqref{ex-5pt}:
\be
\Phi_{11}^{\mathsf{I}}\, \partial_2 \kn \;\ibp\, - \kn\, \partial_2  \Phi_{11}^{\mathsf{I}} = 0,
\en
since $\Phi_{11}^{\mathsf{I}}$ does not depend on $z_2$. For the remaining terms in \eqref{ex-5pt}, we can add the second row to the first one to get:
\begin{align}
\setlength{\arraycolsep}{5pt}
\begin{vmatrix}
	\Phi_{11}^{\mathsf{R}} & \partial_1 \kn \\
	\Phi_{12} & \partial_2 \kn \\
\end{vmatrix} = \begin{vmatrix}
\Phi_{11}^{\mathsf{R}} + \Phi_{12} & \left( \partial_1 + \partial_2 \right) \kn \\
\Phi_{12} & \partial_2 \kn \\
\end{vmatrix} = \begin{vmatrix}
0 & \left( \partial_1 + \partial_2 \right) \kn \\
\;\Phi_{12} & \partial_2 \kn \\
\end{vmatrix} &\,=\, - \Phi_{12} \left( \partial_1 + \partial_2 \right) \kn\tr
  &\ibp  \kn \left( \partial_1 + \partial_2 \right) \Phi_{12} = 0,
\end{align}
where we have used IBP twice in $z_1$ and $z_2$, and the final line vanishes since $(\partial_1 + \partial_2)$ is symmetric, while $\Phi_{12}$ is antisymmetric in $z_1$ and $z_2$.

In the case for $n=6$ and $r=3, s=4$ we have:
\be\label{n=6eg}
\setlength{\arraycolsep}{5pt}
\det \widetilde{\Phi} = \begin{vmatrix} 
	\Phi^{\sf R}_{11}+\Phi^{\sf I}_{11}    & \Phi_{12}  & \partial_{1}\kn\\ 
	\Phi_{12} & \Phi^{\sf R}_{22}+\Phi^{\sf I}_{22} &  \partial_{2}\kn\\ 
	\Phi_{13} & \Phi_{23}  &  \partial_{3}\kn\\ 
\end{vmatrix},
\en
where $\Phi^{\sf R}_{11} = - \Phi_{12} - \Phi_{13}$, $\Phi^{\sf I}_{11} = - \Phi_{14} - \Phi_{15}$ and similarly $\Phi^{\sf R}_{22} = -\Phi_{21} - \Phi_{23}, \Phi^{\sf I}_{22} = -\Phi_{24} - \Phi_{25}$. The terms proportional to $\Phi_{11}^{\sf I}$ are:
\be\label{ex-6pt}
\setlength{\arraycolsep}{5pt}
\Phi^{\sf I}_{11}\begin{vmatrix}
	\Phi^{\sf R}_{22}+\Phi^{\sf I}_{22} &  \partial_{2}\kn\\
	\Phi_{23}  &  \partial_{3}\kn\\ 
\end{vmatrix}.
\en
Note that this matrix has the same structure as \eqref{ex-5pt} with additional terms inside $\Phi_{22}^{\sf R}$ and $\Phi_{22}^{\sf I}$ that do not affect the argument used for showing \eqref{ex-5pt} integrates to zero. Here, terms involving $\Phi_{22}^{\sf I}$ once again do not contribute since $\Phi^{\sf I}_{11} \Phi^{\sf I}_{22} \partial_3 \kn$ vanishes using IBP for $z_3$. Similarly, the remaining terms in \eqref{ex-6pt} vanish:
\begin{align}
\setlength{\arraycolsep}{5pt}
\Phi^{\sf I}_{11}\begin{vmatrix}
	\Phi^{\sf R}_{22} &  \partial_{2}\kn\\
	\Phi_{23}  &  \partial_{3}\kn\\ 
\end{vmatrix} = \Phi^{\sf I}_{11}\begin{vmatrix}
\Phi^{\sf R}_{22} + \Phi_{23} &  (\partial_{2} + \partial_{3})\kn\\
\Phi_{23}  &  \partial_{3}\kn\\ 
\end{vmatrix} &\,=\, \Phi^{\sf I}_{11} \left( - \Phi_{21} \partial_3 \kn - \Phi_{23} (\partial_2 + \partial_3) \kn \right)\tr
&\ibp  \kn  \left( \partial_3 ( \Phi^{\sf I}_{11} \Phi_{21}) + (\partial_2 + \partial_3) \Phi_{23} \right) = 0,
\end{align}
where the first term vanishes as it is independent of $z_3$, and the second term vanishes by symmetry. We conclude that the term \eqref{ex-6pt} does not contribute to \eqref{n=6eg}. Similarly, all terms proportional to $\Phi_{22}^{I}$ vanish in \eqref{n=6eg}, so we have are left with:
\begin{align}
\setlength{\arraycolsep}{5pt}
\begin{vmatrix} 
	\Phi^{\sf R}_{11}    & \Phi_{12}  & \partial_{1}\kn\\ 
	\Phi_{12} & \Phi^{\sf R}_{22} &  \partial_{2}\kn\\ 
	\Phi_{13} & \Phi_{23}  &  \partial_{3}\kn\\ 
\end{vmatrix} = \begin{vmatrix} 
0    & 0  & (\partial_{1}+\partial_{2}+\partial_{3})\kn\\ 
\Phi_{12} & \Phi^{\sf R}_{22} &  \partial_{2}\kn\\ 
\Phi_{13} & \Phi_{23}  &  \partial_{3}\kn\\ 
\end{vmatrix} &\,=\, \begin{vmatrix} 
\Phi_{12} & \Phi^{\sf R}_{22} \\ 
\Phi_{13} & \Phi_{23}\\ 
\end{vmatrix} (\partial_{1}+\partial_{2}+\partial_{3})\kn \tr
&\ibp \kn\, (\partial_{1}+\partial_{2}+\partial_{3}) \begin{vmatrix} 
\Phi_{12} & \Phi^{\sf R}_{22} \\ 
\Phi_{13} & \Phi_{23}\\ 
\end{vmatrix} = 0,
\end{align}
which once again is zero since the final determinant is totally antisymmetric in $z_1, z_2, z_3$.

Having illustrated the algorithm on examples, the general strategy is now clear. The irrelevant terms can be dropped from the determinant by a recursive elimination, since they reduce to $\Phi_{aa}^{\sf I}$ times a lower-point matrix modified with terms that are independent of $\partial_b \kn$ for all rows $b$ in this lower-point matrix. The resulting matrix $\widetilde{\Phi}$ consists only of the relevant terms, which cancel out by symmetry. More precisely, since the all the elements in any ${\bf v}_a$ sum to zero, we add all the bottom rows to the first one without changing the value of the determinant, which then becomes up to a sign:
\be
\det \widetilde{\Phi} = \det \Phi^{[1,i,j,n]}_{[r,s,s+1,n]} \sum_{\substack{a=1\\ a \neq i,j,n}}^{n}\partial_a \kn.
\en
Applying integration by parts for variable $z_a$ in each term in the sum gives:
\be\label{final}
\det \widetilde{\Phi} = - \kn \sum_{\substack{a=1\\ a \neq i,j,n}}^{n-3} \partial_a \det \Phi^{[1,i,j,n]}_{[r,s,s+1,n]}.
\en
Since $\det \Phi^{[1,i,j,n]}_{[r,s,s+1,n]}$ is a polynomial in the arguments $s_{bc} / \sigma_{bc} z_{bc}$ for $b,c \in \{1,2,\ldots,n\} - \{i,j,n\}$, the expression \eqref{final} gives zero due to the identity:
\be
\sum_{\substack{a=1\\ a \neq i,j,n}}^{n}\partial_a \left( \frac{s_{bc}}{\sigma_{bc} z_{bc}} \right) = \frac{s_{bc}}{\sigma_{bc}} \left( \partial_b + \partial_c \right) \frac{1}{z_{bc}} = 0.
\en
We conclude that $\det \widetilde{\Phi} \ibp 0$, which is equivalent to showing the claim that \eqref{detPhi} is independent of the choice of labels $i,j,r,s$.

\subsection{Independence of the Reference Punctures $\sigma_q, z_q$}

The only dependence on the reference punctures $\sigma_q$ and $z_q$ comes from the diagonal terms in the matrix $\Phi(\sigma,z)$. They can be rewritten as:
\be\label{phiaa}
\Phi_{aa}=-\sum_{\substack{c=1 \\ c\neq a}}^n s_{ac} \left( \frac{1}{\sigma_{qa}}+\frac{1}{\sigma_{ac}} \right) \left( \frac{1}{z_{qa}}+\frac{1}{z_{ac}} \right),
\en
it is straightforward to check the independence of $z_q$ already on the level of the matrix entries, before taking the determinant and performing integration. Thus we can set $z_q \to \infty$ and concentrate on proving the independence of $\sigma_q$. 

For this purpose let us write \eqref{phiaa} as:
\be\label{phiaa2}
\Phi_{aa} \,=\, -\sum_{\substack{c=1 \\ c\neq a}}^n \frac{s_{ac}}{\sigma_{ac} z_{ac}}+\frac{\partial_a  \log \kn}{\sigma_{aq}}.
\en
where the sum is independent of $\sigma_q$. We will refer to $\Phi_{aa}$ as a \emph{diagonal term}, no matter where it appears in the matrix $\Phi$ after removal of columns and rows. We want to show that
\be\label{a2}
\kn\; \frac{\partial}{\partial \sigma_q} {\det}^\prime  \Phi(\sigma,z) \,\ibp\, 0.
\en
Let us pick $i,j,r,s,k$ and use $\varphi:=\Phi^{[ijk]}_{[rsk]}$ that will simplify the notation. The left hand side of \eqref{a2} then reads:
\be\label{expandfactor}
 \frac{(-1)^{i+j+r+s}}{\sigma_{rs}\sigma_{ik}\sigma_{jk}\, z_{ij} z_{rk}z_{sk}}\kn \frac{\partial}{\partial \sigma_q}  \det \varphi,
\en
and we have:
\be
\kn\frac{\partial}{\partial \sigma_q}  \det \varphi \,=\, \kn \frac{\partial}{\partial \sigma_q}\sum_{\alpha \in S_{n-3}}\mathrm{sgn}(\alpha)\prod_{a=1}^{n-3}\varphi_{a, \alpha_a}.
\en
Defining the set $\mathcal{D}_\alpha=\{a \,|\, 1\leq a \leq n-3\textrm{ such that }\varphi_{a,\alpha_a}\textrm{ is a diagonal term}\}$, one separates the products in the above sum into two parts:
\be\label{sep}
\kn\frac{\partial}{\partial \sigma_q}  \det \varphi \,=\, \kn\!\! \sum_{\alpha\in S_{n-3}}\mathrm{sgn}(\alpha) \prod_{a\notin \mathcal{D}_\alpha}\varphi_{a, \alpha_a} \;\frac{\partial}{\partial \sigma_q}\prod_{a\in \mathcal{D}_\alpha}\varphi_{a, \alpha_a}.
\en
Translating back into the matrix $\Phi$, let us define another set $\mathcal{D'_\alpha}=\{b \,|\, \textrm{\;if\;} \exists\, a\in \mathcal{D}_\alpha \mathrm{\;such\; that\;}  \Phi_{bb}=\varphi_{a,\alpha_a}\}$. Therefore, the important part we need to consider is:
\bes
\kn \frac{\partial}{\partial \sigma_q}\prod_{a\in \mathcal{D}_\alpha}\varphi_{a, \alpha_a}&=&\;
\kn \frac{\partial}{\partial \sigma_q}\prod_{b\in \mathcal{D'_\alpha}}\Phi_{bb}\tr
&=& \;\kn\; \frac{\partial}{\partial \sigma_q} \prod_{b\in \mathcal{D'_\alpha}} \left( \sum_{\substack{c=1\\c \neq b}}^{n} \frac{s_{bc}}{\sigma_{bc} z_{bc}} + \frac{\partial_b  \log \kn}{\sigma_{bq}} \right)\tr
&=& \;\kn \sum_{\lambda\in \mathcal{D'_\alpha}}\left( -\frac{\partial_\lambda  \log \kn}{\sigma_{\lambda q}^2}\prod_{\substack{b\in \mathcal{D'_\alpha}\\  b\neq \lambda}} \left( \sum_{\substack{c=1\\c \neq b}}^{n} \frac{s_{bc}}{\sigma_{bc} z_{bc}} + \frac{\partial_b \log \kn}{\sigma_{bq}} \right) \right)\tr
&\ibp&\;\kn \sum_{\substack{\lambda,m\in \mathcal{D'_\alpha}\\ \lambda\neq m }} \left( \frac{s_{\lambda m}}{\sigma_{\lambda q}\sigma_{mq}\sigma_{\lambda m}z_{\lambda m}^2 }\prod_{\substack{a\in \mathcal{D'_\alpha}\\a\neq \lambda,m }}\left( \sum_{\substack{c=1\\c \neq a}}^{n} \frac{s_{ac}}{\sigma_{ac} z_{ac}} +\frac{\partial_a  \log \kn}{\sigma_{aq}} \right)\right)=0.\qquad\qquad
\ens
We have used integration by parts for each index in $\mathcal{D'_\alpha}$ separately. Since none of the additional terms $\prod_{a\notin \mathcal{D}_\alpha}\varphi_{a, \alpha_a}$ from \eqref{sep} depends on $z_\lambda$ for $\lambda \in \mathcal{D'_\alpha}$, they do not contribute. In the final step the summand of $\lambda$ and $m$ is antisymmetric under $\lambda \leftrightarrow m$ and hence the whole sum vanishes, proving the required result.

\subsection{Independence of the Choice of Label $k$}

So far we have shown that for fixed $k$, the expression \eqref{detPhi} is invariant of the choice of $i,j,r,s,\sigma_q,z_q$, forming an equivalence class $[i,j,r,s,\sigma_q,z_q]_k$. It remains to prove that it is also independent of the choice of $k$, or in other words that two equivalence classes $[i,j,r,s,\sigma_q,z_q]_k$ and $[i,j,r,s,\sigma_q,z_q]_{k'}$ intersect for $k \neq k'$. Consider a representative $[k',j,k',s,\sigma_q,z_q]_k$ of the first class. It is equal to a representative $[k,j,k,s,\sigma_q,z_q]_{k'}$ of the second class, and therefore they intersect. It follows that \eqref{detPhi} is independent of the choice of $k$. This concludes the proof that \eqref{detPhi} is a well-defined object.

\section{\label{sec:BuildingBlockB}Abelianization of the String Parke--Taylor Factor}

In this appendix we study the procedure called \emph{abelianization}, which has been recently utilized in the context of Z-theory \cite{Mafra:2016mcc,Carrasco:2016ygv}. It is a way of stripping away colour degrees of freedom from a subset of particles. Let us leave the first $r$ particles untouched, and abelianize the remaining $n-r$. The way to do it is to replace all the colour factors in the second set with identity. The abelianized partial amplitudes are then coefficients of traces with $r$ generators, e.g., the partial amplitude with identity ordering $\I_r$ is proportional to ${\rm Tr}(T^{a_1} T^{a_2} \cdots T^{a_r})$.

We can repeat this procedure on the string Parke--Taylor. Formally, we have
\bes
\PTap(\I_r | \times_{n-r}) = \int_{D(\I_r)\times {\mathbb R}^{n-r}}\!\!\!\!\!\!\! \KN\; {\det}^\prime \Phi(\sigma, z).
\ens
In practice, we can calculate this object with the methods of \cite{Carrasco:2016ygv}. Let us illustrate it with an example for $n$ even and all particles abelianized, in which case the answer will take an interesting form. Using the results from \cite{Carrasco:2016ygv} with a different normalization, we have
\bes
\PTap(\times_n) &=& \left(\frac{2}{\pi \alpha'} \right)^{n-2}\!\!\!\! \sum_{\beta \in S_{n-1}} \PTap(1,\beta(2,3,\ldots,n))\tr
&=& \left(\frac{2i}{\pi \alpha'} \right)^{n-2}\!\!\!\! \sum_{\beta \in S_{n-2}} \PTap(1,\beta(2,3,\ldots,n-1),n)  \prod_{i=2}^{n-1} \sin\left( \pi \alpha' k_{\beta(i)} \cdot (k_1 + \cdots + k_{\beta(i-1)}) \right),\nonumber
\ens
where the sine factors arise from monodromy of the disk integrals. For instance, for $n=4$ we obtain
\bes
\PTap(\times_4) &=& (\Pfp \mathsf{A}_4)^2 \left( - \frac{2}{\pi^3} \Gamma(\alpha' s) \Gamma(\alpha' t) \Gamma(\alpha' u)  \Big(\! \sin (\pi \alpha' s) + \sin (\pi \alpha' t) + \sin (\pi \alpha' u) \Big) \right)\tr
&=& (\Pfp \mathsf{A}_4)^2 \left( 1 + \frac{1}{4}\alpha'^2 \zeta_2 (s^2 + t^2 + u^2)  - \alpha'^3 \zeta_3\, s t u + \frac{3}{40} \alpha'^4 \zeta_2^2 (s^2 + t^2 + u^2)^2 + \dots \right).\nonumber
\ens
The leading order cancels due to the $U(1)$ decoupling identity, and the expansion begins at ${\cal O}(\alpha'^2)$. In general, monodromy factors conspire to cancel all the terms up to ${\cal O}(\alpha'^{n-2})$. The leading order organizes itself into a square of a reduced Pfaffian \cite{Cachazo:2014xea} of an antisymmetric matrix $\mathsf{A}$, given by $\mathsf{A}_{ab} = k_a \cdot k_b/\sigma_{ab}$.

The expressions for higher number of particles are quite lengthly. However, we find that the sub-leading part can be succinctly written using a new building block,
\bes\label{PTAbelianized}
\PTap(\times_n) &=& (\Pfp \mathsf{A}_n)^2 + \alpha'^2 \zeta_2\, {\cal B}_n + \ldots.
\ens
The definition of ${\cal B}_n$ can be understood as a \emph{generalized dimensional reduction} \cite{Cachazo:2014xea} of the object ${\cal P}_n$ calculating the ${\rm Tr} F^3$ amplitudes \cite{He:2016iqi,Zhang:2016rzb}. Such object is only non-vanishing for the case when $n$ is even, in agreement with $\PTap(\times_n)$, which has the same property.

We first define an auxiliary term,
\bes
{\mathsf S}_{(\alpha)} = \frac{k_{\alpha(1)} \cdot k_{\alpha(2)}}{\sigma_{\alpha(1), \alpha(2)}} \frac{k_{\alpha(2)} \cdot k_{\alpha(3)}}{\sigma_{\alpha(2), \alpha(3)}} \cdots \frac{k_{\alpha(m)} \cdot k_{\alpha(1)}}{\sigma_{\alpha(m), \alpha(1)}},
\ens
where $m$ is the length of the permutation $\alpha$. It can be used to construct permutation invariants,
\bes
P({\mathsf S})_{i_1, i_2, \ldots, i_m} = \!\!\!\!\!\! \sum_{\substack{|I_1|=i_1,\,\ldots,\, |I_r|=i_m\\ \text{all even}}}\!\!\!\!\!\! {\mathsf S}_{(I_1)} {\mathsf S}_{(I_2)} \cdots {\mathsf S}_{(I_m)}.
\ens
Here, the sum proceeds over all cyclically-inequivalent permutations of even lengths $i_1$,$i_2$,$\ldots$,$i_m$, for example:
\bes
P({\mathsf S})_{2,2} &=& {\mathsf S}_{(1,2)}{\mathsf S}_{(3,4)} + {\mathsf S}_{(1,3)}{\mathsf S}_{(2,4)} + {\mathsf S}_{(1,4)}{\mathsf S}_{(2,3)},\tr
P({\mathsf S})_{4} &=& {\mathsf S}_{(1,2,3,4)} + {\mathsf S}_{(1,2,4,3)} + {\mathsf S}_{(1,3,2,4)} + {\mathsf S}_{(1,3,4,2)} + {\mathsf S}_{(1,4,2,3)} + {\mathsf S}_{(1,4,3,2)}.
\ens
Using these invariants, we can construct a determinant of the matrix ${\mathsf A}_n$ \cite{Cachazo:2014xea} as a sum over all even partitions of the set $(1,2,\ldots,n)$ weighted with minus signs,
\bes
\det {\mathsf A}_n =\!\!\! \sum_{\substack{i_1 \leq \cdots \leq i_m \, \text{even}\\ i_1 + \ldots + i_m = n}}\!\! (-1)^{n-m}\, P({\mathsf S})_{i_1,i_2,\ldots,i_m} = 0.
\ens
However, it is known \cite{Cachazo:2014xea} that this matrix has two null vectors and hence its determinant is zero. We know that the object ${\cal B}_n$ we are looking for has to have the same mass dimension, and also be permutation invariant. In order to make a non-vanishing definition, following \cite{He:2016iqi} we introduce another weight in the sum, 
\be
{\cal B}_n =\!\!\! \sum_{\substack{i_1 \leq \cdots \leq i_m\, \text{even}\\ i_1 + \ldots + i_m = n}}\!\! (-1)^{n-m}\,m\, P({\mathsf S})_{i_1,i_2,\ldots,i_m}.
\en
Explicitly, we have:
\bes
{\cal B}_4 &=& 2 P({\mathsf S})_{2,2} - P({\mathsf S})_{4},\tr
{\cal B}_6 &=& -3 P({\mathsf S})_{2,2,2} + 2P({\mathsf S})_{2,4} - P({\mathsf S})_{6},\tr
{\cal B}_8 &=& 4 P({\mathsf S})_{2,2,2,2} - 3 P({\mathsf S})_{2,2,4} + 2 P({\mathsf S})_{2,6} + 2 P({\mathsf S})_{4,4} - P({\mathsf S})_{8}.
\ens
We have checked numerically that these definitions plugged into \eqref{PTAbelianized} reproduce the abelianized Z-theory amplitudes \cite{Carrasco:2016ldy} up to $n=8$. For example, the $n=6$ case reproduces the result \cite{Carrasco:2016ldy}:
\bes
Z_\times (\I_6) \bigg|_{\alpha'^3} \!\!\!\!\!&=& \frac{\pi^2}{12}\Bigg( -\frac{(s_{12} + s_{23})(s_{12}^2 + s_{12} s_{23} + s_{23}^2)(s_{45}+s_{56})}{s_{123}} + 4 s_{12} s_{23} s_{234} + 4 s_{12} s_{23} s_{345} - 4 s_{12} s_{23} s_{34}\tr
&&\qquad\quad + 2 s_{12} s_{23} s_{56} + 2 s_{12} s_{23} s_{45} + 2 s_{12} s_{34} s_{123} + 2 s_{12} s_{34} s_{234} + s_{12} s_{34} s_{345} + s_{12}^3 + 2 s_{12}^2 s_{45} \tr
&&\qquad\quad + 2 s_{12}^2 s_{234} - 2 s_{12} s_{234}^2 - 4 s_{12} s_{123} s_{234} - 2 s_{23} s_{123} s_{234} - 4 s_{34} s_{123} s_{234} -  s_{12} s_{45} s_{123}/2\tr
&&\qquad\quad -  s_{12} s_{45} s_{345}/2 + s_{123}^2 s_{234} + s_{123}s_{234}^2 +  s_{12} s_{34} s_{56}/3 + 4 s_{123} s_{234} s_{345}/3 +\; {\rm cyclic}\;\Bigg).\nonumber
\ens

\bibliographystyle{JHEP}
\bibliography{references}

\end{document}